# Studies of Parton Propagation and Hadron Formation in the Space-Time Domain

W. K. Brooks[a,d] and H. Hakobyan[b,c,d]

[a]*Departamento de Física y Centro de Estudios Subatómicos,*
*Universidad Técnica Federico Santa María,*
*Avda. España 1680, Casilla 110-V, Valparaíso, Chile.*
[b]*Department of Physics, Yerevan State University,*
*1 Alex Manoogian St, 375049 Yerevan, Armenia.*
[c]*Department of Physics, Yerevan Physics Institute,*
*2 Alikhanyan Brothers St.,375036 Yerevan, Armenia.*
[d]*Physics Division, Thomas Jefferson National Accelerator Facility*
*12000 Jefferson Avenue, Newport News, Virginia 23606.*

**Abstract.** Over the past decade, new data from HERMES, Jefferson Lab, Fermilab, and RHIC that connect to parton propagation and hadron formation have become available. Semi-inclusive DIS on nuclei, the Drell-Yan reaction, and heavy-ion collisions all bring different kinds of information on parton propagation within a medium, while the most direct information on hadron formation comes from the DIS data. Over the next decade one can hope to begin to understand these data within a unified picture. We briefly survey the most relevant data and the common elements of the physics picture, then highlight the new Jefferson Lab data, and close with a prospective for the future**.**

## INTRODUCTION

Measurements of space-time characteristics of QCD are now becoming feasible for the first time using atomic nuclei as fempto-scale systems to analyze short distance processes. The propagation of partons is a topic of interest to multiple communities, including common features spanning the RHIC and LHC heavy ion programs, Drell-Yan measurements at Fermilab, the HERMES experiment at DESY, and Jefferson Lab. Two characteristic times are now accessible to measurement, as well as potentially the quark energy loss in passing through strongly interacting systems, and a new understanding of hadronization mechanisms.

The characteristic times accessible in deep inelastic scattering are illustrated in Fig. 1. In this picture, the struck quark following a hard interaction propagates for some distance, radiating gluons. If the path of propagation includes distance through a strongly-interacting medium, the gluon radiation is enhanced proportional to the gluon density present, and the component of the momentum transverse to the initial direction is increased on average. Following the gluon-emission stage, a color-singlet system is formed, ending the first stage of the reaction and beginning the next. Once a full hadron is formed, the second stage is complete. The characteristic time of the first stage is referred to as the production time, $\tau_p$, and that of the second stage is the formation time, $^h\tau_f$. While the production time may be approximately independent of the hadron species formed for light quarks, the formation time must depend on the hadron characteristics such as its final radius.

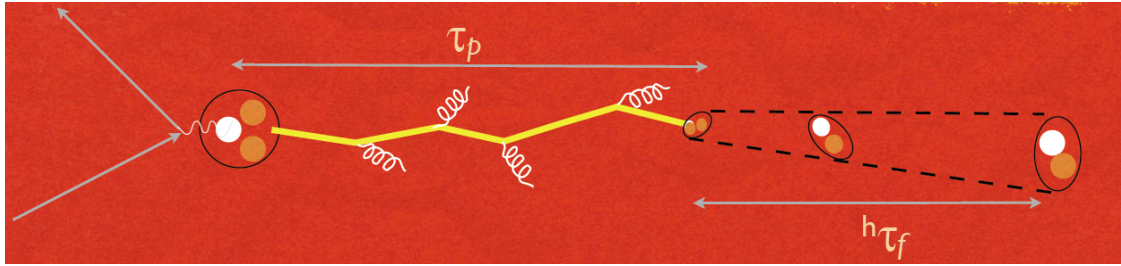

**FIGURE 1.** A schematic diagram illustrating the characteristic times associated with color confinement for the case of deep inelastic scattering. The production time $\tau_p$ is the average lifetime of the propagating colored quark. During this phase of propagation, the quark emits gluons, thus losing energy and providing color flux available for production of subsequent hadrons. The rate of medium-stimulated gluon emission is expected to be slightly greater than that which occurs in vacuum, causing an increase of the width of the transverse momentum distribution of the final state hadron ("transverse momentum broadening") compared to the width seen for the same process in vacuum. The hadron formation time $^h\tau_f$ depends on the hadron type. This stage begins when gluon emission ceases and a color singlet state is formed and ends when a full-sized hadron has been formed.

## $P_T$ BROADENING

Transverse momentum broadening for an observed final state hadron is defined as $\Delta p_T^2 = \langle p_T^2 \rangle_A - \langle p_T^2 \rangle_D$ where the transverse direction is defined relative to the average initial quark direction and the mean transverse momentum squared of the hadrons from deuterium nucleus "D" is subtracted from that for larger nucleus "A." This procedure partially compensates for the intrinsic quark momentum and the isospin averaging for the larger nucleus. This quantity can be accessed in semi-exclusive deep inelastic scattering and in the Drell-Yan reaction.

Fig. 2 gives a naïve picture of what might be expected for a quark propagating through a nucleus. As can be seen from this picture, if the production length $\tau_p$ is short, then the amount of transverse momentum broadening $\Delta p_T^2$ is expected to be small overall, and not depend significantly on the nuclear radius. If, on the other hand, $\tau_p$ is much longer than nuclear dimensions, then the broadening is expected to be proportional to the nuclear radius. In this case it is also, to first order, independent of $\tau_p$. In the intermediate case where $\tau_p$ is longer than, e.g. the diameter of carbon, but shorter than the diameter of lead, a non-linear dependence results. This dependence can be unfolded to extract $\tau_p$. Thus, this fundamental characteristic time is accessible to experiment, but only within a window of energies: enough energy to enter the DIS kinematics, but not so much that the process is time-dilated beyond nuclear dimensions.

Precise data have recently been obtained from the HERMES experiment at DESY and from CLAS at Jefferson Lab. An overview of these preliminary data is shown in

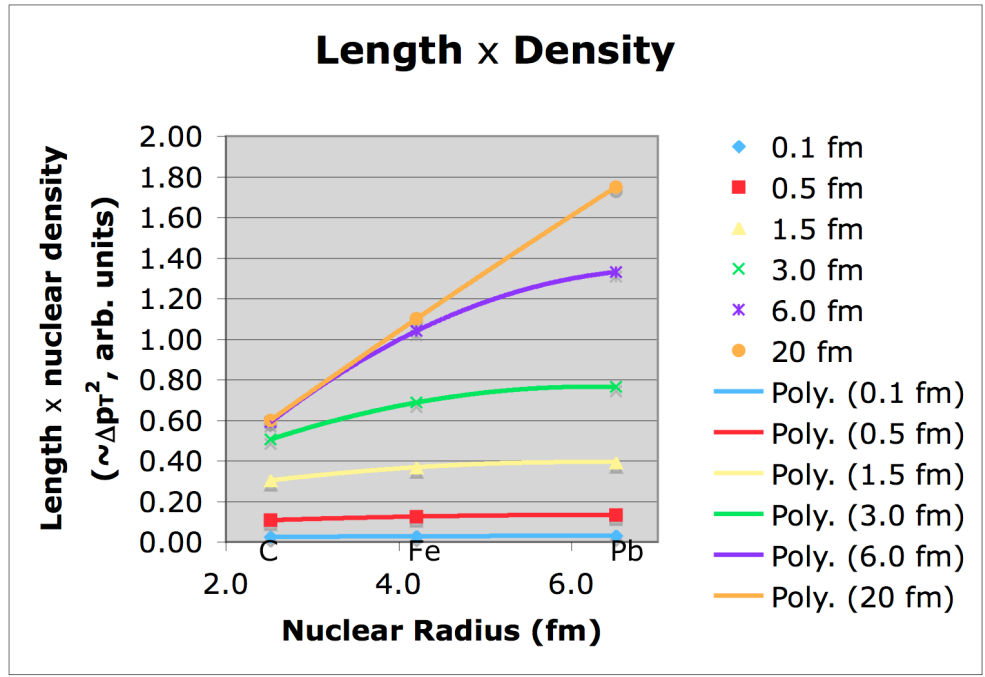

**FIGURE 2.** A plot of the expected behaviors for transverse momentum broadening due to geometrical effects in a simple classical model. The broadening is expected to be proportional to the product (production length) × (nuclear density); the plot shows the numerical average of this quantity integrated over realistic nuclear density distributions for carbon, iron, and lead nuclei. For the shortest lengths, e.g., 0.1 fm, the broadening is small and does not depend strongly on nuclear radius. For the longest production lengths, the broadening is proportional to the nuclear radius and independent of the production length. Thus, there is no sensitivity for extraction of production lengths which are significantly longer than nuclear dimensions, such as measurements at high energies.

Fig. 3, along with data from the Drell-Yan reaction. This overview plot is intended to display global features of the data: it includes several particle types and a number of bins in $\nu$, $Q^2$, and z for the DIS data. In addition, the energy scales of the measurements are substantially different: 800 GeV protons for D-Y, 27 GeV positrons for HERMES, and 5 GeV electrons for JLab.

The disparity between the different measurements in Fig. 3 suggests the possibility of an energy dependence, which has been noted previously in a different connection, see below. However, if this is the explanation, it is not a monotonic dependence, since the lowest energy data do not have the smallest amount of broadening. The effective values of $\langle z \rangle$ for all the measurements in the plot are somewhat different, which may shift the vertical scale differently for different datasets. Another known difference between the JLab and HERMES data is the coverage in Bjorken x. The HERMES data extend down to approximately 0.02, while the JLab data are above 0.1. This can be significant since for x<0.1, quark pair production becomes a significant mechanism

[1], potentially reducing the amount of broadening [2]. An additional possible effect that could explain the difference is additional sources of broadening that come in at the lower energies, such as elastic scattering of the prehadron. Simple estimates of this effect suggest it is negligible, however, because the cross section rises at low energies, a careful calculation may be merited. Thus, while these known effects would help to explain the situation to some extent, the difference between the JLab and HERMES broadening cannot said to be fully understood yet. It is to be hoped that more insight will come from the eventual analysis of the multidimensional data, such as that shown in Fig. 4.

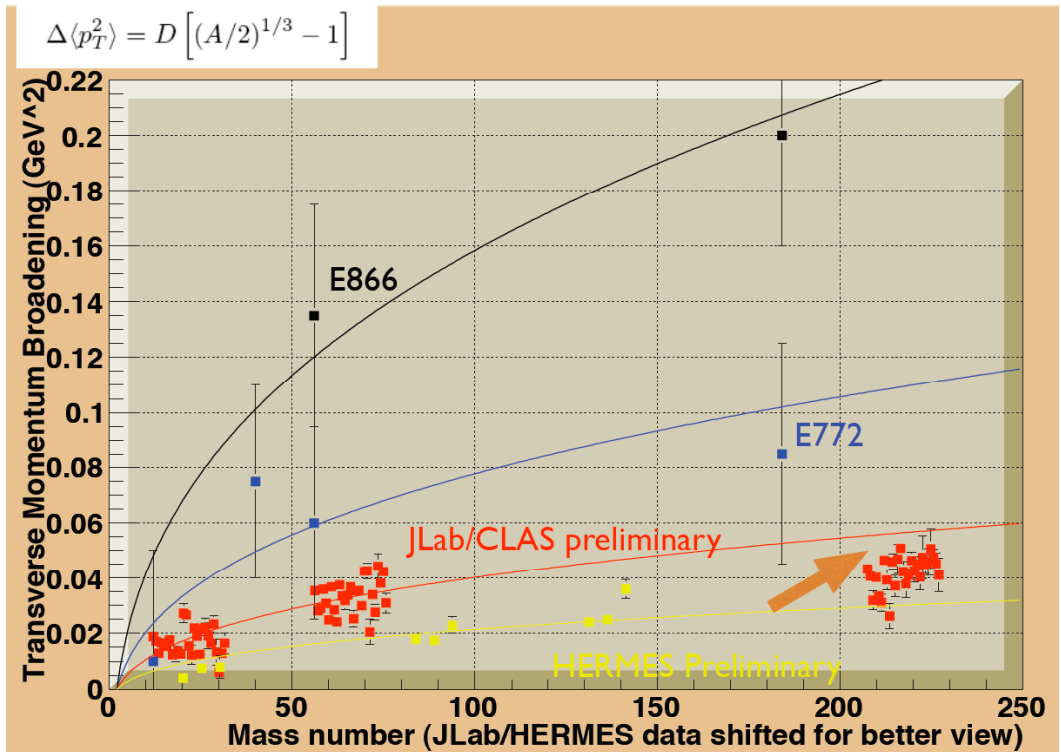

**FIGURE 3.** A plot of the new, high-precision DIS data together with the older Drell-Yan data. The plots associated with the bottom curve are the preliminary HERMES data, those of the second curve from the bottom are the preliminary CLAS data, and the upper two data sets are from older Drell-Yan data. For the HERMES and Jefferson Lab data, the points are arbitrarily shifted rightward so that the error bars are visible. The JLab data are for positive pions in 27 three-dimensional bins in $\nu$, $Q^2$, and z. The HERMES data are for positive and negative pions and for positive kaons. The curves follow the equation in the upper left corner, with the parameter D adjusted to match the data. The JLab data for the heaviest nucleus (lead) fall below the curve, suggesting that there is sensitivity to measure the production length from these data (see caption for Fig. 2).

A framework that has proven to be useful at higher energies is the color dipole model [3], which is able to describe H1/Zeus data over a wide range in x and $Q^2$ [4] and which has been shown to be connected to transverse momentum broadening of a propagating quark. This work yields the simple relation $\Delta p_T^2 = 2C\rho_A L$ where $\rho_A$ is the nuclear density, L is the path length through the medium, and $C = C(r_T, s)$ is a slowly-changing function of the center of mass energy squared $s$, and the transverse separation distance of a color dipole $r_T$; C is closely connected to the gluon density for small $r_T$. One can either assume the value of $C$ is known from higher energy studies, and derive an average production length, or use the data from several nuclei and extract $C$ at the energy of the measurement. Studies of this type have been performed [5], finding production times in excess of 5 fm/c for 4 GeV quarks (ν = 4 GeV) under specific assumptions.

There have also been calculations for $\Delta p_T^2$ within a perturbative QCD framework. Specific predictions for $\Delta p_T^2$ for positive pions were published by Qiu [6]. In this work it was implicitly assumed that the production length was longer than the nuclear dimensions.

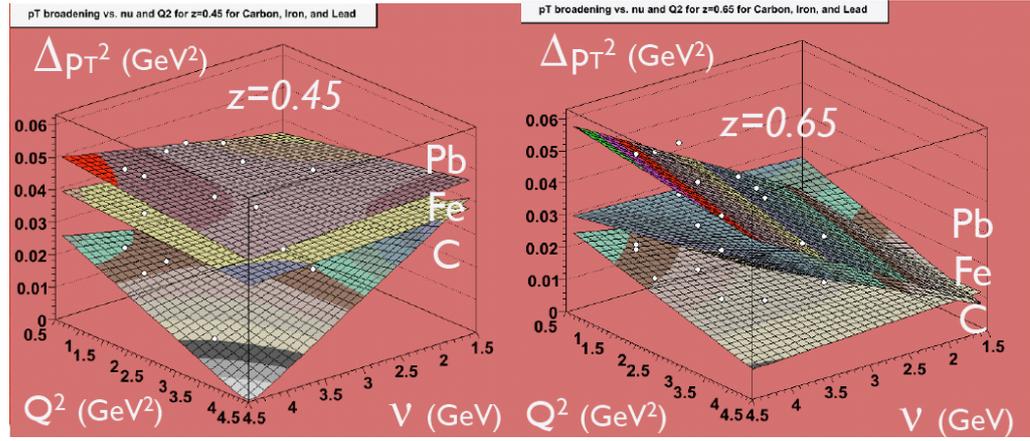

**FIGURE 4.** Examples of fits of the CLAS transverse momentum broadening in three-dimensional bins (ν, $Q^2$, and z) for three nuclei. Analyses of this kind can be used to extract the production length as a function of three variables.

A quark-gluon correlation function model was used to estimate the broadening, with a strength parameter λ:

$$T_{qF}^A(x,Q^2) = \lambda^2 A^{1/3} q^A(x,Q^2)$$

where $q$ is the effective twist-2 quark distribution of a nucleus normalized by the atomic weight $A$. Within this approach, the strength parameter can now be fixed using the new precise data. Thus, these data may provide a direct measure of a dominant quark-gluon correlation function within the nucleus.

Other studies connecting parton propagation to pQCD are compiled in [7]. Note that the precision data of Fig. 3 give a value of the transport coefficient $\langle p_T^2 \rangle / \langle L \rangle$ that is smaller than any of the values quoted there, even taking into account the $z^2$ scaling factor [5] connecting quark and hadron $p_T^2$.

Planned extensions to the studies presented here are feasible with the existing JLab dataset. These take the form of extracting further information from associated particles accompanying the reactions, such as bremsstrahlung photons and low-energy protons.

The bremsstrahlung photons that are emitted by the propagating quark were noted by [1] and detailed calculations have been performed by [8]. The latter calculations suggest that there are two physical processes which can interfere: emission of the photon at the point of the hard interaction, and emission during the multiple scattering through the medium. Observation of these photons would be a further validation of the physical picture, and would permit the determination of the two parameters needed within the pQCD picture for a complete first-order description of the multiple scattering process.

It has been suggested that observation of slow protons in DIS can be used to study the target fragmentation region [9]. These calculations suggest that sensitivity to discriminate between a spectator-type behavior and target fragmentation can be obtained by the production cross section for protons moving in the backward direction. From the CLAS data, these could also be studied in coincidence with the fast forward-moving current fragment.

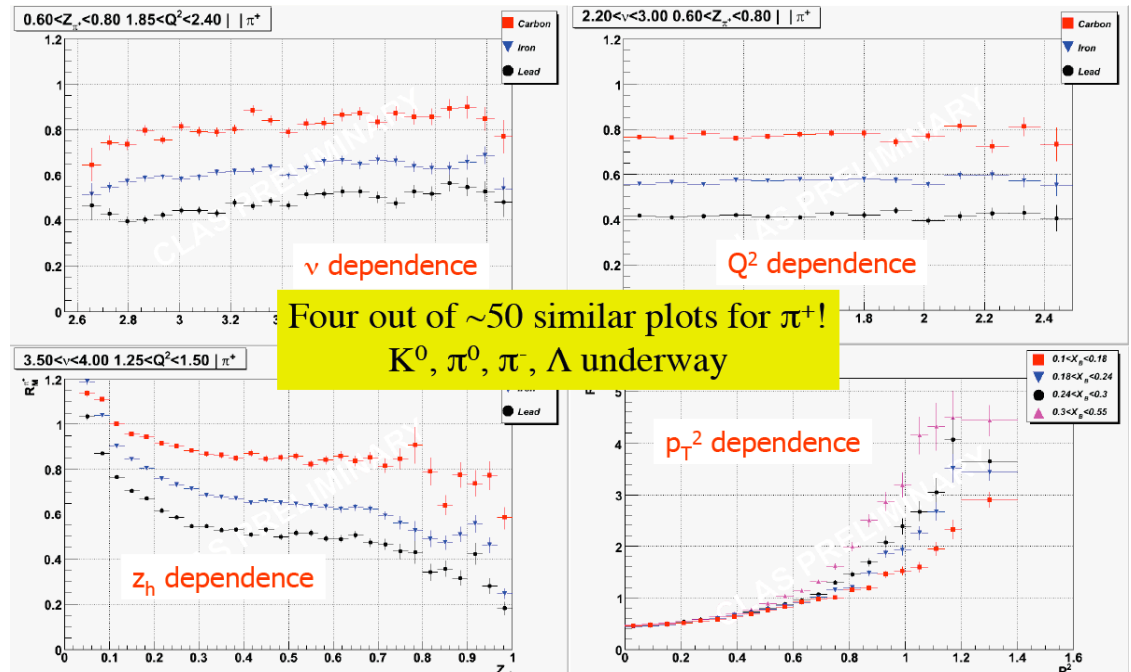

**FIGURE 5.** Examples of multivariable measurements of the hadronic attenuation ratio for positive pions in CLAS (preliminary data). Each point is binned in three kinematic variables for each nucleus. For the plots of the $\nu$, $Q^2$, and $z$ dependences, the uppermost data is for carbon and the lowest is for lead. In the plot of the transverse momentum dependence (lower right panel), the upper curve is for the highest values of Bjorken x and the lower curves are for increasingly lower values of Bjorken x, down to x=0.1.

## HADRON ATTENUATION

Hadron attenuation refers to the apparent modification of the fragmentation functions in nuclei relative to deuterium or hydrogen. For most of the kinematic range it is a reduction of produced hadron flux, although at low $z$ and high $p_T$ it is an enhancement. The typical variable used to quantify the modification is the "hadronic multiplicity ratio" $R$, defined as:

$$R(v,Q^2,z,p_T,\phi) = \frac{\left[\frac{N_h(v,Q^2,z,p_T,\phi)}{N_e(v,Q^2)}\right]_{A,DIS}}{\left[\frac{N_h(v,Q^2,z,p_T,\phi)}{N_e(v,Q^2)}\right]_{D,DIS}}$$

(where $\phi$ is the angle between the lepton scattering plane and the virtual photon-hadron plane). HERMES has published multiplicity ratios for a variety of targets from helium to xenon [10]. These data are very important because for the first time they were able to measure the multiplicity ratio for identified hadrons: $\pi^{+,-,0}$, $K^{+,-}$, proton and antiproton. This is a critical step in the process of determining hadronic formation times, in particular the dependence on, e.g., mass and flavor for baryons and mesons.

| hadron | $c\tau$ | mass (GeV) | flavor content | detection channel | Production rate per 1k DIS event |
|---|---|---|---|---|---|
| $\pi^0$ | 25 nm | 0.13 | $u\bar{u}d\bar{d}$ | $\gamma\gamma$ | 1100 |
| $\pi^+$ | 7.8 m | 0.14 | $u\bar{d}$ | direct | 1000 |
| $\pi^-$ | 7.8 m | 0.14 | $d\bar{u}$ | direct | 1000 |
| $\eta$ | 0.17 nm | 0.55 | $u\bar{u}d\bar{d}s\bar{s}$ | $\gamma\gamma$ | 120 |
| $\omega$ | 23 fm | 0.78 | $u\bar{u}d\bar{d}s\bar{s}$ | $\pi^+\pi^-\pi^0$ | 170 |
| $\eta'$ | 0.98 pm | 0.96 | $u\bar{u}d\bar{d}s\bar{s}$ | $\pi^+\pi^-\eta$ | 27 |
| $\phi$ | 44 fm | 1.0 | $u\bar{u}d\bar{d}s\bar{s}$ | $K^+K^-$ | 0.8 |
| $f1$ | 8 fm | 1.3 | $u\bar{u}d\bar{d}s\bar{s}$ | $\pi\pi\pi\pi$ | - |
| $K^+$ | 3.7 m | 0.49 | $u\bar{s}$ | direct | 75 |
| $K^-$ | 3.7 m | 0.49 | $\bar{u}s$ | direct | 25 |
| $K^0$ | 27 mm | 0.50 | $d\bar{s}$ | $\pi^+\pi^-$ | 42 |
| $p$ | stable | 0.94 | $ud$ | direct | 530 |
| $\bar{p}$ | stable | 0.94 | $\bar{u}\bar{d}$ | direct | 3 |
| $\Lambda$ | 79 mm | 1.1 | $uds$ | $p\pi^-$ | 72 |
| $\Lambda(1520)$ | 13 fm | 1.5 | $uds$ | $p\pi^-$ | - |
| $\Sigma^+$ | 24 mm | 1.2 | $us$ | $p\pi^0$ | 6 |
| $\Sigma^0$ | 22 pm | 1.2 | $uds$ | $\Lambda\gamma$ | 11 |
| $\Xi^0$ | 87 mm | 1.3 | $us$ | $\Lambda\pi^0$ | 0.6 |
| $\Xi^-$ | 49 mm | 1.3 | $ds$ | $\Lambda\pi^-$ | 0.9 |

**FIGURE 6.** Examples of hadrons accessible to CLAS12 following the 12 GeV Upgrade of JLab. The hadrons have been selected to be stable on a distance scale corresponding to nuclear sizes, and to be measurable by CLAS12. Production rates were estimated with Pythia. These hadrons can be used for extraction of hadronic formation lengths, including meson and baryon mass and flavor dependence.

The HERMES measurements are limited to one-dimensional binning for most of the data, but they span a wide range in z and v.

The preliminary JLab data have a much more limited range in $v$ and $z$, but have a statistical sample that is two orders of magnitude greater than HERMES. As a result, three-dimensional kinematic bins can be formed for each target nucleus, so that the multidimensional dependences of $R$ can be studied. Examples of these distributions are given in Fig. 5 for positive pions. From these datasets, detailed models of hadron formation can be tested and refined.

These studies may have relevance for the heavy ion programs at RHIC and LHC. A more detailed understanding of the timescales and mechanisms of hadron formation could influence the interpretation of the heavy ion data. Beyond the obvious contributions to the last stages of the collision process, there have been suggestions that this physics could play a more prominent role in jet suppression than is generally believed [11].

## FUTURE PROSPECTS

In the near future, in addition to further analysis of the JLab data for associated particle production as already discussed, there is a new Drell-Yan experiment planned to run at Fermilab, experiment E906. This run, with 120 GeV proton beam on nuclear targets, includes many improvements over the previous experiments, and the lower beam energy is intended to reduce theoretical ambiguities in the analysis.

The Jefferson Lab 12 GeV Upgrade will follow a few years later, bringing the capabilities of higher electron beam energies and more luminosity. Experimental Halls B and D will feature large acceptance spectrometers dedicated to electron beams of $10^{35}$ cm$^{-2}$s$^{-1}$ and tagged, collimated photon beams of $10^8$ photons per second, respectively. The Hall D capabilities will allow dedicated study of the properties of the prehadron, particularly in diffractive kinematics, while the Hall B capabilities will feature studies in deep inelastic kinematics. A list of the hadrons accessible is shown in Fig. 6, together with their mass and flavor characteristics. In addition to a dedicated program for vector meson transparency in the nuclear medium [12], a broad program of measurements is planned for JLab experiment E12-06-117 [13] which will provide a wealth of new information on hadronic formation lengths.